\documentclass[11pt,twoside,a4paper]{article}
\usepackage{myartmods}
\usepackage[utf8]{inputenc}
\usepackage{datetime}\usdate
\usepackage[english]{babel}

\usepackage{amsmath}
\usepackage{amsfonts}
\usepackage{amssymb}

\usepackage{graphicx}
\usepackage[caption=false]{subfig}
\usepackage{tabularx}
\usepackage{booktabs}
\usepackage{authblk}
\usepackage[colorlinks]{hyperref}

\numberwithin{equation}{section}
\numberwithin{figure}{section}
\numberwithin{table}{section}

\catcode`@=11
\def\dref{d^{\text{ref}}}
\def\vref{v^{\text{ref}}}
\newcommand{\drefk}[1]{d_#1^{\text{ref}}}
\newcommand{\vrefk}[1]{v_#1^{\text{ref}}}
\def\minim{\mathop{\operator@font minimize}}
\def\minimize#1{{\displaystyle\minim_{#1}}}
\def\minimizetwo#1#2{{\displaystyle\minim_{\substack{#1\\#2}}}}
\def\maxim{\mathop{\operator@font maximize}}
\def\maximize#1{{\displaystyle\maxim_{#1}}}
\def\subject{\mathop{\operator@font subject\ to}}
\def\diag{\mathop{\operator@font diag}}
\catcode`@=12

\begin{document}
\title      {Explicit optimization of plan quality measures in intensity-modulated 
             radiation therapy treatment planning} 
\author[1,2]{Lovisa Engberg\thanks{Corresponding author: loven140@kth.se}}
\author[2]  {Kjell Eriksson}
\author[1]  {Anders Forsgren}
\author[2]  {Bj\"{o}rn H\aa rdemark}
\affil[1]   {Optimization and Systems Theory, Department of Mathematics, KTH Royal Institute of 
             Technology, Stockholm SE-100 44, Sweden}
\affil[2]   {RaySearch Laboratories, Sveav\"{a}gen 44, Stockholm SE-103 65, Sweden}
\date       {Manuscript \\ January 24, 2017}
\markboth   {Explicit IMRT treatment plan optimization}{}
\maketitle\thispagestyle{empty}

\begin{abstract} 
Conventional planning objectives in optimization of intensity-modulated radiotherapy treatment (IMRT) plans are designed to minimize the violation of dose-volume histogram (DVH) thresholds using penalty functions. Although successful in guiding the DVH curve towards these thresholds, conventional planning objectives offer limited control of the individual points on the DVH curve (doses-at-volume) used to evaluate plan quality. In this study, we abandon the usual penalty-function framework and propose planning objectives that more explicitly relate to DVH statistics. The proposed planning objectives are based on mean-tail-dose, resulting in convex optimization. We also demonstrate how to adapt a standard optimization method to the proposed formulation in order to obtain a substantial reduction in computational cost.

We investigate the potential of the proposed planning objectives as tools for optimizing DVH statistics through juxtaposition with the conventional planning objectives on two patient cases. Sets of treatment plans with differently balanced planning objectives are generated using either the proposed or the conventional approach. Dominance in the sense of better distributed doses-at-volume is observed in plans optimized within the proposed framework, indicating that the DVH statistics are better optimized and more efficiently balanced using the proposed planning objectives. 

\smallskip
\noindent{\bf Keywords:} Planning objectives, convex optimization, mean-tail-dose
\end{abstract}

\section{Introduction}
Conventional treatment planning in intensity-modulated radiation therapy (IMRT) is often described as a time-consuming trial-and-error process, as it requires the repeated solution of successively fine-tuned treatment plan optimization problems before requirements on plan quality are met. The need for re-optimization partly stems from a methodological difference between the mathematical planning objectives and the clinical evaluation criteria. While conventional planning objectives minimize the violation of dose statistics thresholds using (quadratic) penalty functions, attention is rarely paid during plan quality assessment to the optimal level of deviation. What rather influences plan quality is the actual dose statistics of the dose distribution. 

The vast amount of resources spent on treatment planning has motivated the development of strategies for automated approaches that require limited user guidance. A particular interest has recently grown in knowledge-based planning, where achievable yet desirable dose statistics thresholds are predicted with machine-learning techniques ahead of optimization. The expected outcome is a less need for fine-tuning. Several methods have been proposed in this direction \cite{Appenzoller2012,Shiraishi2016,SkarpmanMunter2015,Song2015,Wu2009,Wu2011} and their success in producing high-quality treatment plans while reducing user interaction has been investigated in subsequent evaluations \cite{Chang2016,Fogliata2015,Tol2015}. Another suggested approach is computerization of the typical trial-and-error scheme adopted by treatment planners. While some of these methods are site-specific \cite{Zhang2011}, others are developed with the ambition to handle a large class of cases \cite{Gintz2016,Hazell2016,Xhaferllari2013}. 

Still aiming for a less demanding planning process, we return to the underlying inconsistency between mathematical planning objectives and clinical evaluation criteria. We abandon the usual penalty-function framework and suggest planning objectives that more explicitly relate to plan quality measures, in our case to dose-volume histogram (DVH) statistics. The proposed planning objectives are based on the wellknown concept of mean-tail-dose often referred to by the financial counterpart conditional value-at-risk (CVaR). The use of CVaR-based objectives to obtain qualified approximations of optimal value-at-risk (corresponding to DVH statistics) is a frequent tool in finance \cite{Alexander2006,Rockafellar1997} since this leads to convex optimizaion. As for treatment planning, mean-tail-doses were originally incorporated as constraints of a penalty-function based formulation by Romeijn \emph{et al.} \cite{Romeijn2006}. The aim was to give a tractable alternative to bounds on the mathematically intractable DVH statistics. In following studies, mean-tail-doses have been used in, e.g., lexicographic optimization of prostate plans \cite{Clark2008} and as constraints in optimization of brachytherapy \cite{Holm2013}. The present study investigates their merit as tools for optimizing the DVH statistics to be assessed in the plan evaluation process.

Leaving the penalty-function framework has the attractive effect that dose statistics thresholds are no longer required, neither is therefore the process of fine-tuning them. Nevertheless, a need remains to find the objective weights that give the desired tradeoff between conflicting evaluation criteria. Multicriteria optimization (MCO) techniques, by which the procedure of balancing planning objectives can be largely shortened, have been developed for the conventional case through rigorous research (see, e.g., \cite{Bokrantz2013a,Bokrantz2015,Craft2013,Fredriksson2013} for recent advances). Although beyond the scope of this study, we envisage that similar MCO techniques can be as advantageously combined with the proposed planning objectives. The outcome would then be a planning process that has little need for parameter tweaking.

The paper is organized as follows. In Section~\ref{sec:Formulation}, the conventional and proposed formulations of planning objectives are presented. Section~\ref{sec:CompStud} provides the set-up of and results from a computational study where the two formulations are compared. In Section~\ref{sec:Method}, a method is described to handle the large-scale dimensions of the proposed optimization formulation. Existence of such a method is of utmost importance for the clinical applicability of the proposed approach; however, this section can be omitted by the reader who is only interested in the outcome of the method.

\section{Formulation of planning objectives}\label{sec:Formulation}
Treatment plan MCO is generally formulated 
\begin{equation*}\label{eq:genMCO}
\begin{aligned}
	& \minimize{x \in \mathcal{X}} && \big[\,f_1(x) \cdots f_K(x)\,\big]^T,
\end{aligned}
\end{equation*}
where each of the $K$ planning objectives $f_1,\ldots,f_K$ somehow relates to a plan evaluation criterion, and where $\mathcal{X}$ is the set of feasible treatment variables. We limit this study to fluence map optimization (FMO), so that $\mathcal{X}$ equals the convex set $\{x \in \mathbb{R}^n: x \geq 0\}$ of physically realizable fluence maps of $n$ beamlets. However, the theory and methods applied permit the more general assumption that $\mathcal{X}$ be any convex polytope of treatment variables that relate to voxel dose through a linear mapping. In addition to FMO, that assumption is valid for, e.g., certain formulations of sliding window optimization and spot weight optimization in proton therapy.

Treatment plan MCO formulations are usually solved under the notion of Pareto optimality. A Pareto optimal plan is produced, among other techniques, by solving a weighted-sum instance where the $K$ planning objectives have been aggregated into one using nonnegative objective weights. Each combination of weights results in a different Pareto optimal plan, and all possible Pareto optimal plans form the Pareto set. See \cite{Bokrantz2013b} for a thorough introduction to MCO with focus on treatment planning.

This study concerns plan evaluation criteria relating to the cumulative dose-volume histogram (DVH). As a function of the fluence map $x$, the DVH statistics quantifies the least dose received by the hottest volume fraction $\vref$ of the region-of-interest (ROI) $r$. A mathematical definition is 
\begin{equation*}
	D(x;\,r,\vref) = 
	\min \{ d \in \mathbb{R} : \sum_{\substack{j \in V_r:\\p_j^Tx \geq d}} \Delta^r_j \leq \vref\},
\end{equation*}
where $V_r$ collects the voxels of ROI $r$, $\Delta_j^r$ is the relative volume of $r$ located in voxel $j$, and $p_j^T$ is the $j$th row of the dose deposition coefficient matrix 
so that $p_j^Tx$ is the voxel dose received by voxel $j$. The compact notation $D(x)$ is 
used throughout, and $D(x)$ is referred to as a dose-at-volume. A mathematical drawback of the DVH statistics is its non-convexity and non-differentiability, making it intractable for optimization purposes.

\subsection{Conventional planning objectives}
\begin{figure}[t]\centering
\subfloat{\centering\includegraphics[width=160pt]{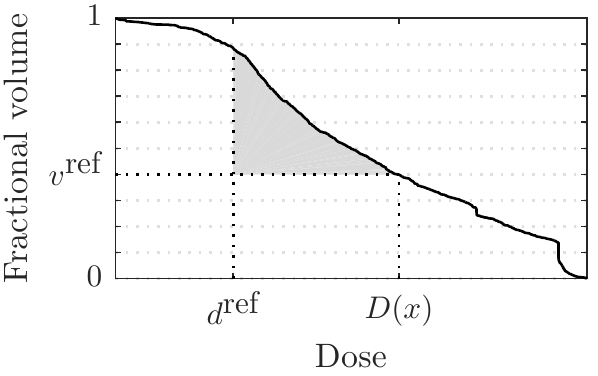}}\\
\subfloat{\centering\includegraphics[width=160pt]{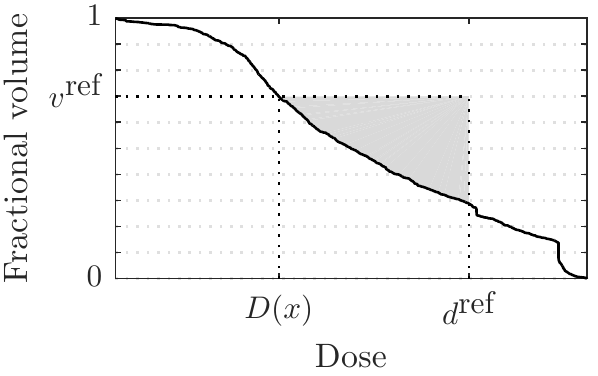}}
\caption{Voxel overdoses (top) and underdoses (bottom) (shaded) from the threshold $\dref$ illustrated in the DVH graph. A conventional planning objective $q^+(x)$ or $q^-(x)$ is the sum of the voxel overdoses or underdoses squared so that when minimized, the DVH curve is pushed towards the point $(\dref,\vref)$, aiming to meet a maximum or a minimum DVH criterion at volume fraction $\vref$.\label{fig:ConventionalObjectives}}
\end{figure}

Conventional planning objectives are designed to push the DVH curve of a ROI towards the point $(\dref,\vref)$, with $\dref$ a DVH statistics threshold. Aiming to meet maximum or minimum DVH criteria at volume fraction $\vref$, they quadratically penalize voxel overdose or underdose from $\dref$ while omitting the hottest $\vref$ or coldest $1-\vref$ volume fraction. Thus, by construction, only the voxel doses that fall between $\dref$ and the dose-at-volume $D(x)$ are penalized, as seen in Figure~\ref{fig:ConventionalObjectives}.
The mathematical definitions are 
\begin{equation*}
	q^+(x;\,r,\vref,\dref) = 
		\sum_{\substack{j \in V_r:\\ \dref \leq p_j^Tx \leq D(x)}} \mkern-12mu \Delta_j^r\,\big(p_j^Tx - \dref)^2
\end{equation*}
for penalizing overdose, and 
\begin{equation*}
	q^-(x;\,r,\vref,\dref) = 
		\sum_{\substack{j \in V_r:\\ D(x) \leq p_j^Tx \leq \dref}} \mkern-12mu \Delta_j^r\,\big(\dref - p_j^Tx\big)^2
\end{equation*}
for penalizing underdose. Compact notations $q^+(x)$ and $q^-(x)$ are used throughout.

A conventional planning objective is both non-convex and non-differentiable, yet we want to draw the attention to another problematic aspect: its limited ability to optimize or merely control the actual dose-at-volume $D(x)$. For instance, large gaps between $\dref$ and $D(x)$ can be assigned relatively small penalties. The tool available for the user to reach satisfactory plan quality is to iteratively modify the thresholds and re-optimize the plan. This is a trial-and-error challenge, since the impact of $\dref$ adjustments on the dose-at-volume is unclear. Another strategy, now formalized by MCO techniques, is to choose a utopian $\dref$ (such as 0~Gy for healthy tissue) and then fine-tune the objective weight to increase or decrease the impact of the penalty. Nevertheless, the inconsistency remains between a conventional planning objective and the pointwise dose-at-volume.

\subsection{Proposed planning objectives}
To the extent that plan quality is assessed by dose-at-volume, an idealistic planning objective is, naturally, equalling $D(x)$ itself. Treatment plan MCO would then amount to solving
\begin{equation}\label{eq:desiredMCO}
\begin{aligned}
	& \minimize{x \in \mathcal{X}} 
	           && \mkern-12mu \big[\,D_1(x) \cdots D_{k_+}(x) -\!\! D_{k_++1}(x) \cdots -\!\! D_K(x)\,\big]^T \\
	& \subject && D_k(x) \leq u_k, \quad k = 1,\ldots,k_+, \\
	&          && D_k(x) \geq l_k, \quad k = k_++1,\ldots,K,
\end{aligned}
\end{equation}
where $D_k(x) = D(x; r_k, \vrefk{k})$, and where bounds $l_k$ and $u_k$ restrict $D_k(x)$ to relevant values. Here, the last $K-k_+$ doses-at-volume are subjected to maximization by minimizing the negatives. This idealized formulation is largely intractable: for instance, the inherent non-convexity of $D(x)$ renders \eqref{eq:desiredMCO} a non-convex problem whose global minimum is difficult to find. It forms, however, the basis for the proposed formulation, demonstrating how the usual penalty-function framework is abandoned.

\begin{figure}[t]\centering
\subfloat{\centering\includegraphics[width=160pt]{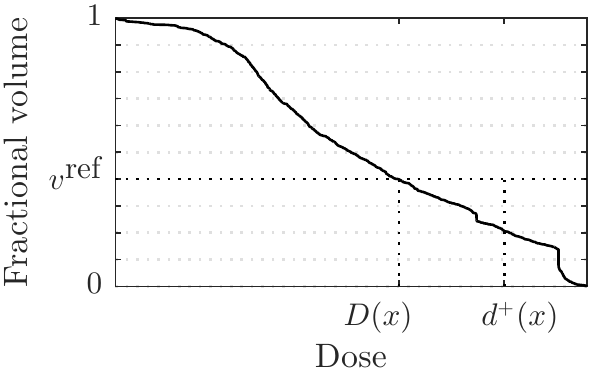}}\\
\subfloat{\centering\includegraphics[width=160pt]{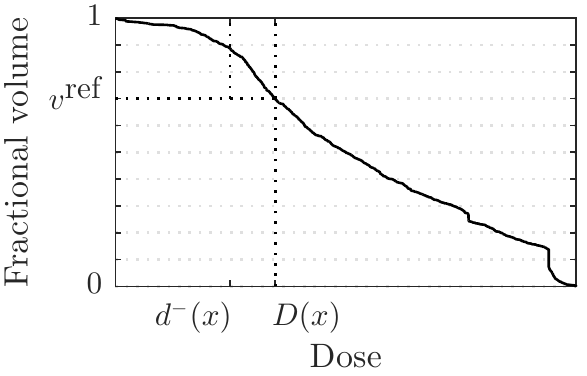}}
\caption{Upper (top) and lower (bottom) mean-tail-doses in relation to dose-at-volume $D(x)$. The upper version $d^+(x)$ is the average of voxel doses greater than $D(x)$ and the lower version $d^-(x)$ is the average of voxel doses less than $D(x)$. When minimizing $d^+(x)$ or maximizing $d^-(x)$, $D(x)$ is pushed to the left or right, respectively.\label{fig:MeanTailDose}}
\end{figure}

We arrive at the proposed formulation by using planning objectives that approximate dose-at-volume by upper or lower mean-tail-dose. The upper or lower version respectively equals the average dose received by the hottest $\vref$ or coldest $1-\vref$ volume fraction, i.e., the average of voxel doses greater or less than $D(x)$ as depicted in Figure~\ref{fig:MeanTailDose}. 
Their mathematical definitions can be established independently of $D(x)$ (see Rockafellar and Uryasev \cite{Rockafellar1997} for the derivation for the financial counterpart CVaR) as
\begin{equation*}\begin{split}
    d^+(x; r, \vref) = & \\
	&\mkern-40mu \min\{\alpha + \frac{1}{\vref}\sum_{j \in V_r} \Delta^r_j \left(p_j^Tx - \alpha\right)_+ : 
					\alpha \in \mathbb{R}\}
\end{split}\end{equation*}
for upper mean-tail-dose, and
\begin{equation*}\begin{split}
	d^-(x; r, \vref) = & \\
	&\mkern-75mu \max\{\alpha - \frac{1}{1-\vref}\sum_{j \in V_r} \Delta^r_j \left(\alpha - p_j^Tx\right)_+ : 
					\alpha \in \mathbb{R}\}
\end{split}\end{equation*}
for lower mean-tail-dose, where $(\,\cdot\,)_+$ denotes the positive part function $\max\{\,\cdot\,, 0\}$. Compact notations $d^+(x)$ and $d^-(x)$ are used throughout. 

There are two favorable aspects of using mean-tail-doses as planning objectives: the convexity of $d^+(x)$ and concavity of $d^-(x)$ (i.e., convexity of the negative of $d^-(x)$) making them suitable for optimization, and the relationship $d^-(x) \leq D(x) \leq d^+(x)$ ensuring that the dose-at-volume is appropriately controlled. Treatment plan MCO with proposed planning objectives becomes 
\begin{equation*}
\begin{aligned}
& \minimize{x \in \mathcal{X}} 
           && \mkern-12mu \big[\,d^+_1(x) \cdots d^+_{k_+}(x) -\!\! d^-_{k_++1}(x) \cdots -\!\! d^-_K(x)\,\big]^T \\
& \subject && d^+_k(x) \leq u_k, \quad k = 1,\ldots,k_+,           \\ 
&          && d^-_k(x) \geq l_k, \quad k = k_++1,\ldots,K,
\end{aligned}
\end{equation*} 
a convex formulation whose Pareto optimal solutions provide pessimistic bounds on the doses-at-volume of the optimized treatment plan. The formulation expands into
\begin{equation}\label{eq:convexFMO}
\begin{aligned}
& \minimizetwo{\alpha_k,d_k \in \mathbb{R},\,x \in \mathcal{X}}{\eta^k \in \mathbb{R}^{m_k}} 
           &&\mkern-12mu \big[\,d_1 \cdots d_{k_+} -\!\! d_{k_++1} \cdots -\!\! d_K\,\big]^T                 \\
& \subject && \alpha_k + \frac{1}{\vrefk{k}}\sum_{j \in V_{r_k}} \Delta^r_j \eta^k_j \leq d_k,    \\
&          && \eta^k_j \geq p_j^Tx - \alpha_k, \enskip \eta^k_j \geq 0, \enskip j \in V_{r_k},    \\[3pt]
&          &&                                                   &&\mkern-120mu k = 1,\ldots,k_+,   \\[8pt]
&          && \alpha_k - \frac{1}{1-\vrefk{k}}\sum_{j \in V_{r_k}} \Delta^r_j \eta^k_j \geq d_k,  \\
&          && \eta^k_j \geq \alpha_k - p_j^Tx, \enskip \eta^k_j \geq 0, \enskip j \in V_{r_k},    \\[3pt]
&          &&                                                   &&\mkern-120mu k = k_++1,\ldots,K, \\[8pt]
&          && d_k \in \big[l_k,u_k\big],                        &&\mkern-120mu k = 1,\ldots,K,
\end{aligned}
\end{equation}
with artificial $m_k$-dimensional variables $\eta^k$ introduced to linearly handle the positive part function, denoting by $m_k$ the number of voxels in ROI $r_k$; and with auxiliary variables $d_k$ introduced for clarity. It should be noted that we have slightly modified the formulation by accepting both upper and lower bounds on $d_k$. The desired (and obtained) effect is removed incentive to minimize or maximize the $k$th mean-tail-dose beyond these limits. In a Pareto optimal solution, $d_k$ takes the value of the $k$th mean-tail-dose or, if any bound on $d_k$ is active, gives a pessimistic bound.

\subsection{A note on planning constraints}
Planning constraints impose bounds on doses-at-volume without providing incentive to exceed the requirements. An upper bound is formulated by conventional means as the constraint $q^+(x) \leq 0$, which corresponds to enforcing the DVH curve to reach the point $(\dref,\vref)$ where thus $\dref$ is the bound. Analogously, a lower bound is imposed by $q^-(x) \leq 0$. In the proposed framework, upper and lower bounds on doses-at-volume are formulated as upper or lower bounds on upper or lower mean-tail-dose, respectively.

\subsection{A note on maximum and minimum dose}
Maximum or minimum doses are conventionally controlled by quadratically penalizing all voxel dose deviations from the threshold $\dref$. The mathematical definitions are
\begin{equation*}
	q^{\text{max}}(x; r, \dref) = \sum_{\substack{j \in V_r:\\ \dref \leq p_j^Tx}} \Delta_j^r\,\big(p_j^Tx - \dref\big)^2
\end{equation*}
for a maximum dose objective, and 
\begin{equation*}
	q^{\text{min}}(x; r, \dref) = \sum_{\substack{j \in V_r:\\ p_j^Tx \leq \dref}} \Delta_j^r\,\big(\dref - p_j^Tx\big)_+^2
\end{equation*}
for a minimum dose objective, coinciding with $q^+(x)$ and $q^-(x)$ for $\vref = 0$ and $\vref = 1$, respectively. The functions are convex and differentiable, but the previous discussion regarding inconsistency with plan quality applies: the planning objectives are successful in pushing the DVH curve towards $\dref$, but have limited ability to control the actual maximum or minimum dose. 

The proposed framework allows direct optimization of the maximum and minimum dose statistics. These are mathematically defined as 
\begin{equation*}
	d^{\text{max}}(x; r) = \min \{ d \in \mathbb{R} : p_j^Tx \leq d, \forall j \in V_r \}
\end{equation*}
for maximum dose, and
\begin{equation*}
	d^{\text{min}}(x; r) = \max \{ d \in \mathbb{R} : p_j^Tx \geq d, \forall j \in V_r \}
\end{equation*}
for minimum dose. Convexity of $d^{\text{max}}$ and concavity of $d^{\text{min}}$ make them suitable for minimization and maximization, respectively, and they can be integrated into \eqref{eq:convexFMO} without changing its characteristics.

The construction of planning constraints for maximum and minimum dose is analogous to the construction of dose-at-volume constraints. For instance, an upper bound on maximum dose is given by $q^{\text{max}}(x; r, \dref) \leq 0$ or $d^{\text{max}}(x; r) \leq u$. It should be noted that these constraints are equivalent if $\dref = u$.

\section{Computational study}\label{sec:CompStud}
The proposed and conventional frameworks are juxtaposed in a preliminary computational study comprising two patient cases. The aim is to get an indication of the relative merit of the planning objectives as tools for optimizing DVH statistics. To this end, we compare the distribution of doses-at-volume among treatment plans generated in either the proposed or the conventional framework. Both patient cases are limited to three planning objectives in order to allow comparison in three-dimensional plots; however, planning constraints are added to increase complexity and clinical relevance.

Weighted-sum instances of the proposed formulation in \eqref{eq:convexFMO} are solved using a MATLAB (MathWorks, Natick, Massachusetts) implementation of the method presented in Section \ref{sec:Method}. The conventional formulation is managed using the SQP solver SNOPT \cite{Gill2005} (Stanford Business Software, Stanford, California). Patient geometries and other problem data, including dose deposition coefficient matrices, are exported from RayStation (RaySearch Laboratories, Stockholm, Sweden).

\subsection{Patient cases}
The cases involve a prostate and a lung cancer patient. The patient geometries are discretized into $74\,877$ and $106\,465$~voxels and the (five coplanar) beams into, in total, $1310$ and $3923$~beamlets, all using 5~mm resolution grids. Planning objectives are chosen to minimize doses-at-volume to respectively the bladder, rectum and entire healthy volume (referred to as the surrounding ROI), and the esophagus, lung and surrounding ROI. The following requirements are added as planning constraints. For the prostate case, a minimum dose of 68~Gy is required to the planning target volume (PTV) and the maximum dose to the entire patient volume (external ROI) must not exceed 72~Gy. For the lung case, a minimum of 66.5~Gy and a maximum of 75~Gy are required to the PTV, and the maximum dose to the surrounding ROI and spinal cord must not exceed 70~Gy and 50~Gy, respectively. All parameters are listed in Table~\ref{tab:ModelParams}. It should be noted that, by restricting to planning constraints for minimum and maximum dose, we ensure that the proposed and conventional formulations define identical feasible regions. 
\begin{table}[t]\centering
\caption{Parameters for planning objectives (o) and constraints (c) used in the prostate and lung case.\label{tab:ModelParams}}
\begin{tabularx}{\columnwidth}{l l l X X X l X X}
\hline\hline
\multicolumn{3}{l}{\bf Prostate case}                                                 \\
&     & \multicolumn{4}{l}{Proposed} & \multicolumn{3}{l}{Conventional}               \\
& ROI &  & $\vrefk{k}$ & $l_k$ & $u_k$ &  & $\vrefk{k}$ & $\drefk{k}$                 \\\hline 
o & Surrounding & $d^+$ &  5\% & 0 & 70 & $q^+$ &  5\% & 0                            \\
o & Bladder     & $d^+$ & 50\% & 0 & 70 & $q^+$ & 50\% & 0                            \\
o & Rectum      & $d^+$ & 20\% & 0 & 70 & $q^+$ & 20\% & 0                            \\[3pt]
c & External & $d^{\text{max}}$ & --- & --- &  72 & $q^{\text{max}}$ & --- & 72       \\
c & PTV      & $d^{\text{min}}$ & --- &  68 & --- & $q^{\text{min}}$ & --- & 68       \\[6pt]
\multicolumn{3}{l}{\bf Lung case}                                                     \\
&     & \multicolumn{4}{l}{Proposed} & \multicolumn{3}{l}{Conventional}               \\
& ROI &  & $\vrefk{k}$ & $l_k$ & $u_k$ &  & $\vrefk{k}$ & $\drefk{k}$                 \\\hline 
o & Surrounding & $d^+$ &  5\% & 0 & 70 & $q^+$ &  5\% & 0                            \\
o & Lung        & $d^+$ & 25\% & 0 & 70 & $q^+$ & 25\% & 0                            \\
o & Esophagus   & $d^+$ & 20\% & 0 & 70 & $q^+$ & 20\% & 0                            \\[3pt]
c & Surrounding & $d^{\text{max}}$ & --- &  --- &  70 & $q^{\text{max}}$ & --- & 70   \\
c & PTV         & $d^{\text{min}}$ & --- & 66.5 & --- & $q^{\text{min}}$ & --- & 66.5 \\
c & PTV         & $d^{\text{max}}$ & --- &  --- &  75 & $q^{\text{max}}$ & --- & 75   \\
c & Spinal cord & $d^{\text{max}}$ & --- &  --- &  50 & $q^{\text{max}}$ & --- & 50   \\
\hline\hline
\end{tabularx}
\end{table}

\subsection{DVH statistics of treatment plan cohorts}
A cohort of differently balanced treatment plans was generated for both patient cases and within both frameworks, resulting in four sets of plans; each treatment plan is the result of a weighted-sum instance. The underlying four sets of objective weight triplets $(w_1,w_2,w_3)$ all included the anchor (one $w_k$ equal to unity and others to zero) and the balanced (all $w_k$ equal) triplets. The remaining combinations varied among the four sets and were sampled from successively refined symmetric grids to have, as far as deemed possible, well-distributed plans. This strategy is primitive in its nature and only convenient for two- or three-dimensional MCO; for higher dimensions, it is recommended to apply techniques similar to those suggested in \cite{Bokrantz2013b}. A total of respectively 55 and 159 plans were generated for the prostate case using the proposed and conventional framework, and 34 and 130 for the lung case.

\begin{figure}[t]\centering
\subfloat{\centering\includegraphics[height=90pt]{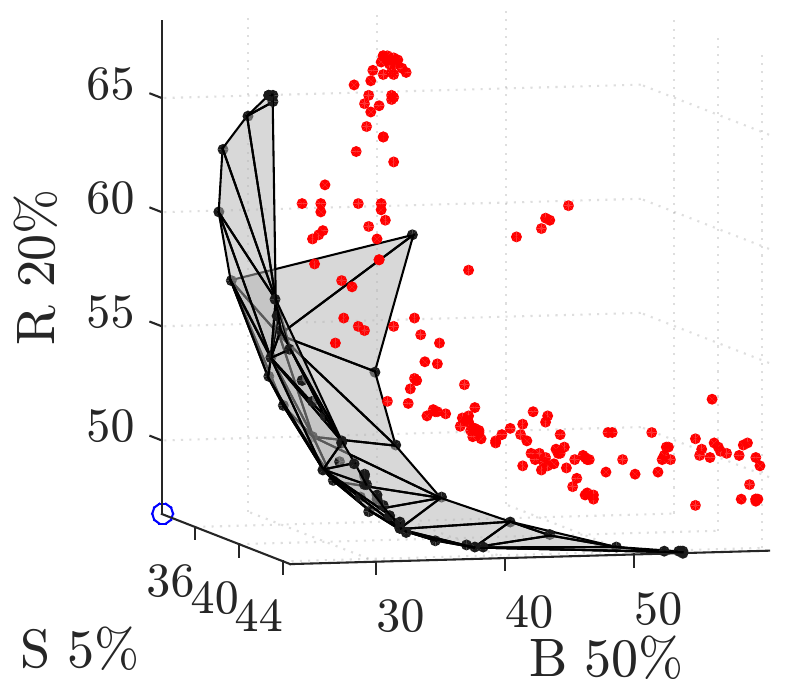}}\hspace*{10pt}
\subfloat{\centering\includegraphics[height=90pt]{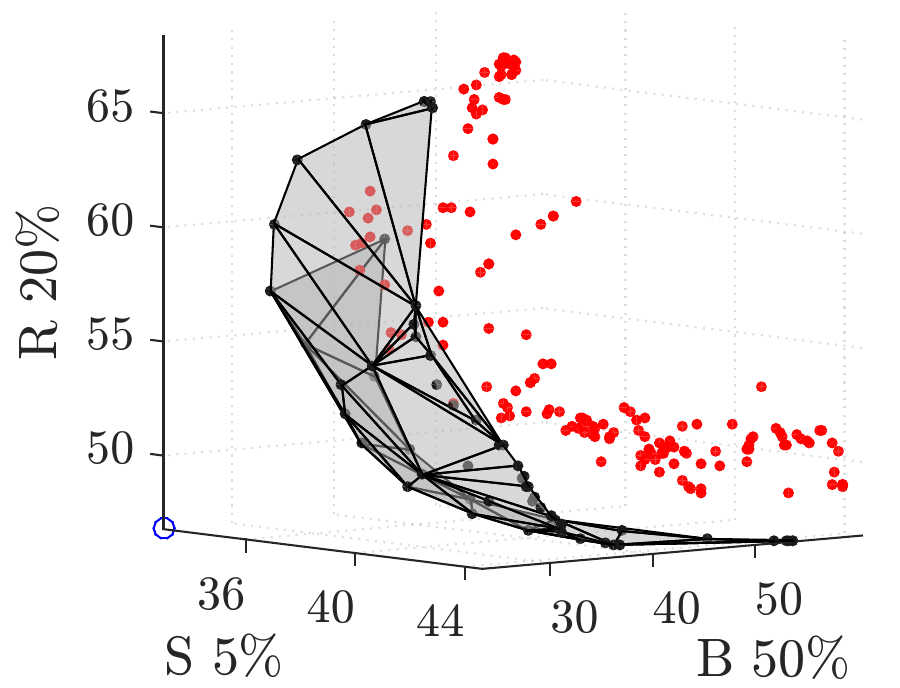}}\hspace*{10pt}
\subfloat{\centering\includegraphics[height=90pt]{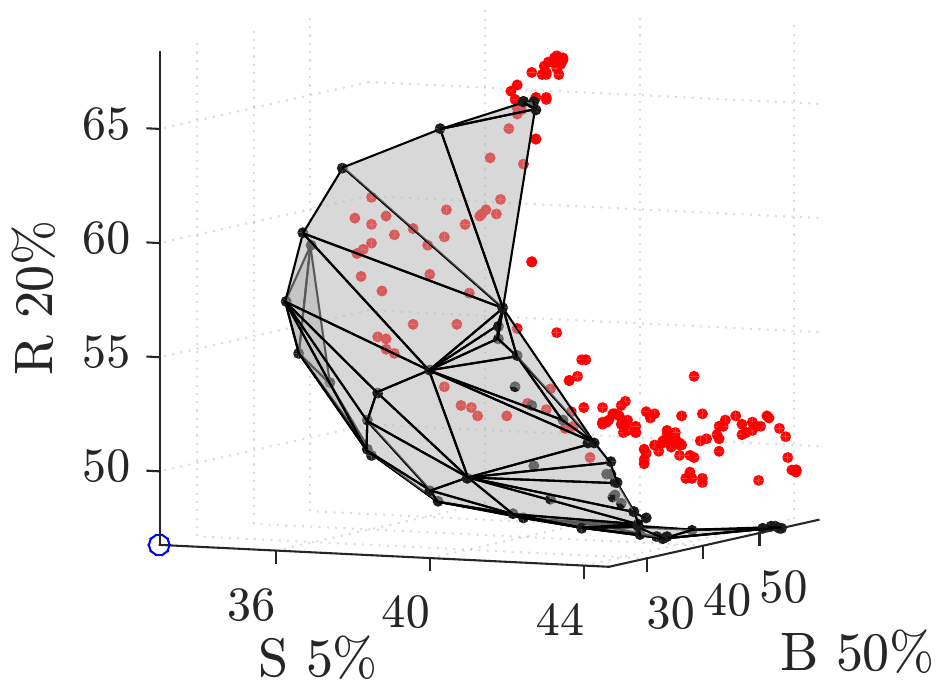}}
\caption{Dose-at-volume triplets [Gy] obtained in prostate plans generated using the proposed (black dots) and conventional (red dots) planning objectives, and the convex hull of the former. The left, middle, and right subfigures show different angles of the plot. The doses-at-volume concern the surrounding ROI (S), bladder (B), and rectum (R). \label{fig:Pareto:prostate}}
\end{figure}

\begin{figure}[t]\centering
\subfloat{\centering\includegraphics[height=90pt]{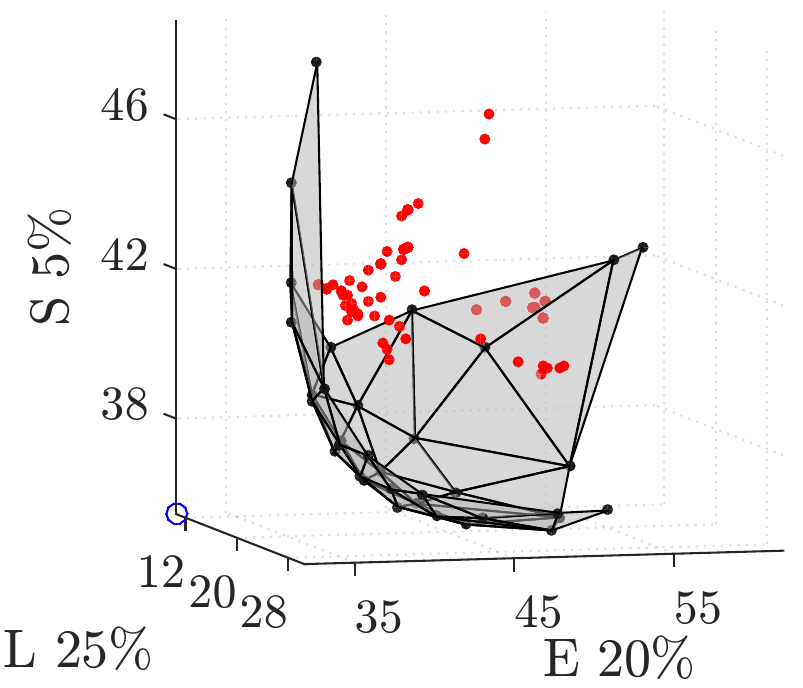}}\hspace*{10pt}
\subfloat{\centering\includegraphics[height=90pt]{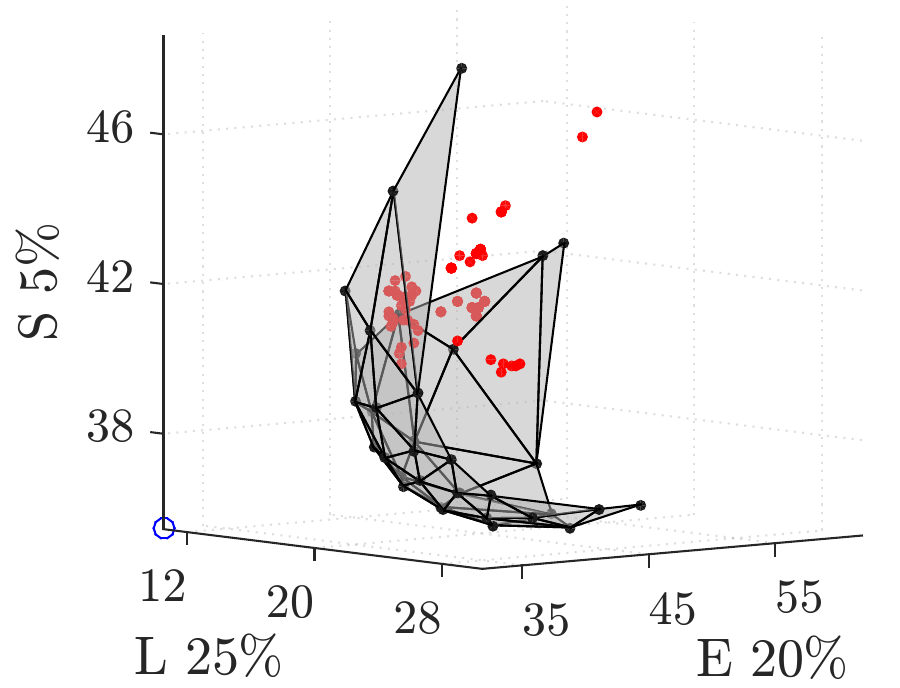}}\hspace*{10pt}
\subfloat{\centering\includegraphics[height=90pt]{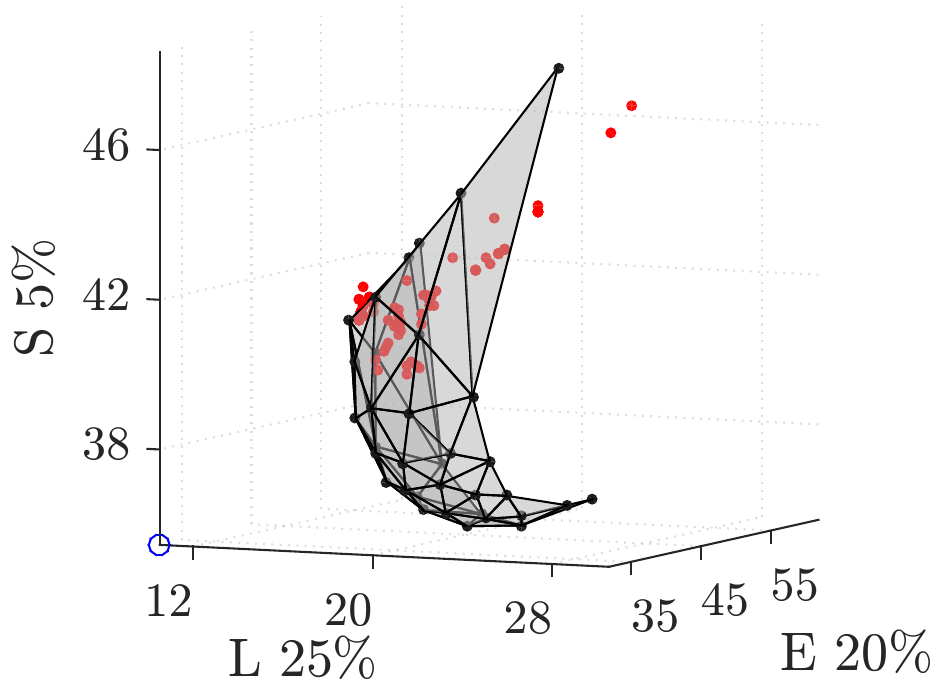}}
\caption{Dose-at-volume triplets [Gy] obtained in lung plans generated using the proposed (black dots) and conventional (red dots) planning objectives, and the convex hull of the former. The left, middle, and right subfigures show different angles of the plot. The doses-at-volume concern the esophagus (E), lung (L), and surrounding ROI (S).\label{fig:Pareto:lung}}
\end{figure}

Each treatment plan was characterized by the three doses-at-volume intended to be optimized. The distribution of such triplets in a three-dimensional coordinate system is illustrated in Figures~\ref{fig:Pareto:prostate}~and~\ref{fig:Pareto:lung}; the left, middle, and right subfigures show different angles to enhance spatial perception and the corner of all-lowest value within the axes limits has been marked with a circle.
The figures give two indications of particular interest. First, the plans optimized using the proposed planning objectives are superior to those generated within the conventional framework in the sense that their convex hull is located closer to the all-lowest point. This dominance is seen for both patient cases and suggests that the proposed framework, despite its approximative nature, provided a more efficient tool for optimizing the DVH statistics. Second, the proposed plans dominate each other to a remarkably smaller extent than the conventional plans. For the lung case in particular, the proposed plans span a wide region in the dose-at-volume domain, as opposed to the conventional plans which give the impression of being randomly scattered. 

A comment is needed regarding the accuracy in terms of fulfilment of planning constraint. While each generated proposed plan strictly satisfied all requirements on minimum and maximum dose, almost all conventional plans violated them. For instance, the median minimum PTV dose of conventional plans admits an underdose of 2.0 Gy in the prostate case and 3.6 Gy in the lung case. Violation within some tolerance is standard; however, SNOPT had difficulties in satisfying the specified tolerance in several tested instances. This property of conventional planning is known, and is briefly discussed and explained in \cite{Fredriksson2012}.

\section{Numerical method}\label{sec:Method}
In this section, the issue of solving weighted-sum instances of \eqref{eq:convexFMO} is discussed. The instances belong to the class of linear programming (LP) problems for which there exists several numerical optimization methods, some commercially available in general-purpose implementations. However, the LP problems of this study require more careful handling due to their size being proportional to the number of voxels. With 5~mm resolution, the number of variables and constraints is easily brought to the order of $10^5$. Below, we describe how the specific structure of \eqref{eq:convexFMO} allows us to eliminate number-of-voxels dependence in systems of linear equations of an interior-point method. The reduced system is proportional to the number of beamlets which, at least for fixed-gantry FMO, usually is orders of magnitude smaller than the number of voxels. 

In demonstrating how problem structure is accounted for, we give a brief description of a standard primal-dual interior-point method. The interested reader is referred to, e.g., \cite{Wright1997} for more thoroughgoing theory. 

\subsection{Interior-point method for specific structure}\label{sec:Method:Specific}
For convenience, we express a weighted-sum instance of \eqref{eq:convexFMO} on the more compact form
\begin{equation}\label{eq:compactForm}
\begin{aligned}
&\minimize{z,\eta} && c_z^Tz + c_{\eta}^T\eta                    \\
&\subject          && A_{zz}    z + A_{z\eta}\eta \geq b_z,      \\
&                  && A_{\eta z}z +          \eta \geq b_{\eta}, \\
&                  && z,\eta                      \geq 0,
\end{aligned}
\end{equation}
where $\eta$ collects vectors $\eta^k$ for all $k$ and $z$ collects the remaining variables. This compact form preserves the key characteristic of \eqref{eq:convexFMO}: the identity coefficient matrix of $\eta$ in the second constraint stemming from constraints $\eta_j^k \geq p_j^Tx-\alpha_k$ and $\eta_j^k \geq \alpha_k-p_j^Tx$. We should also note that the size of $A_{zz}$ is number-of-voxels independent, and that $A_{\eta z}$ is dense due to its rows containing $p_j^T$. 

Associated with the (primal) optimization formulation \eqref{eq:compactForm} is its dual 
\begin{equation*}
\begin{aligned}
&\maximize{y_z,y_{\eta}} && b_z^Ty_z + b_{\eta}^Ty_{\eta}                        \\
&\subject                && A_{zz}^T   y_z + A_{\eta z}^Ty_{\eta} \leq c_z,      \\
&                        && A_{z\eta}^Ty_z +             y_{\eta} \leq c_{\eta}, \\
&                        && y_z,y_{\eta}                          \geq 0.
\end{aligned}
\end{equation*}
An interior-point method applies Newton's method to find a solution to the complementary slackness conditions
\begin{equation*}
\begin{aligned}
y_z^T\!\left(A_{zz}z + A_{z\eta}\eta - b_z\right)         = 0, 
    && \mkern-10mu y_{\eta}^T\!\left(A_{\eta z}z + \eta - b_{\eta}\right) = 0, \\
z^T\!\left(c_z - A_{zz}^Ty_z - A_{\eta z}^Ty_{\eta}\right) = 0, 
    && \mkern-10mu \eta^T\!\left(c_{\eta} - A_{z\eta}^Ty_z - y_{\eta}\right)  = 0,
\end{aligned}
\end{equation*}
while ensuring strict primal and dual feasibility by appropriate step sizes; feasibility and complementarity is necessary and sufficient for optimality. To this end, the left-hand-sides of the conditions are perturbed by some $\mu > 0$ that is successively decreased as the method proceeds. Each iteration hence amounts to solving the system of linear equations
\begin{equation}\label{eq:compactForm:KKTsys}\begin{split}
\arraycolsep=2pt
\left[\begin{array}{ c c | c c }
    D_1                & -ZA_{zz}^T    & 0            & -ZA_{\eta z}^T \\
    Y_zA_{zz}          & D_2           & Y_zA_{z\eta} & 0              \\\hline
    0                  & -HA_{z\eta}^T & D_3          & -H             \\
    Y_{\eta}A_{\eta z} & 0             & Y_{\eta}     & D_4            
\end{array}\right]
\left[\begin{array}{ l }
    \Delta z        \\ 
    \Delta y_z      \\ 
    \Delta \eta     \\ 
    \Delta y_{\eta}
\end{array}\right] = & \\ 
&\mkern-140mu -\left[\begin{array}{ r }
                  Z        D_1 e - \mu e \\
                  Y_z      D_2 e - \mu e \\
                  H        D_3 e - \mu e \\
                  Y_{\eta} D_4 e - \mu e
\end{array}\right]
\end{split}\end{equation}
for some $\mu$, where 
\begin{align*}
& Z = \diag(z),        && D_1 = \diag\left(c_z      - A_{zz}^T   y_z - A_{\eta z}^Ty_{\eta}\right), \\
& Y_z = \diag(y_z),    && D_2 = \diag\left(A_{zz}    z + A_{z\eta}\eta - b_z     \right),           \\
& H = \diag(\eta),     && D_3 = \diag\left(c_{\eta} - A_{z\eta}^Ty_z -             y_{\eta}\right), \\
& Y_{\eta} = \diag(y_{\eta}), && D_4 = \diag\left(A_{\eta z}z +          \eta - b_{\eta}\right).
\end{align*}
The method converges to an optimal solution as the perturbation $\mu$ approaches zero. 

Solving \eqref{eq:compactForm:KKTsys} is a computational challenge since the size of the bottom right block is dependent on the number of voxels. However, accounting for the specific structure of the bottom block permits to solve \eqref{eq:compactForm:KKTsys} in two relatively inexpensive steps: by one solve with the substantially smaller Schur complement 
\begin{equation*}\begin{split}
\arraycolsep=1pt
\left[\begin{array}{ c c }
	D_1 & \!\!\! -ZA_{zz}^T \\ Y_zA_{zz} & D_2
\end{array}\right] - & \\
&\mkern-100mu 
\arraycolsep=1pt
-\left[\begin{array}{ c c }
	0 & \!\!\! -ZA_{\eta z}^T \\ Y_zA_{z \eta} & 0
\end{array}\right]
\arraycolsep=4pt
\left[\begin{array}{ c c }
	D_3 & -H \\ Y_{\eta} & D_4 
\end{array}\right]^{-1}
\arraycolsep=1pt
\left[\begin{array}{ c c }
	0 & \!\!\! -HA_{z \eta}^T \\ Y_{\eta}A_{\eta z} & 0
\end{array}\right]
\end{split}\end{equation*}
whose size is of the same order as the number of beamlets; and by one solve with the bottom block. The computational gain comes from the fact that, as the composite of four diagonal matrices, 
\begin{equation*}
\begin{bmatrix}
	D_3 & -H \\ Y_{\eta} & D_4
\end{bmatrix}^{-1} = \,
\begin{bmatrix}
	M D_4 & M H \\ -MY_{\eta} & MD_3
\end{bmatrix}
\end{equation*}
with $M=\left(D_3 D_4 + Y_{\eta}H \right)^{-1}$ diagonal, both the bottom block and its inverse merely act as the inexpensive operation of scaling and adding two vectors. The result is a dimensionality reduction of several orders of magnitude. 

The main computational cost per iteration now lies in computing the dense matrix product $A_{\eta z}^T D_3 Y_{\eta} A_{\eta z}$ appearing in the Schur complement. A technique is suggested in \cite{Gondzio1996} to accelerate convergence at the expense of multiple solves of \eqref{eq:compactForm:KKTsys} with different right-hand-sides. Each additional solve is relatively inexpensive, since it reuses the Schur complement (and its factorization) of the first solve. A decrease in total computational cost is expected, since the accelerated convergence needs fewer iterations to meet a certain accuracy.

It should be mentioned that interior-point methods require starting points that strictly fulfil inequality constraints. It is therefore convenient to introduce nonnegative slack variables $s_z$, $s_{\eta}$, $\sigma_z$, and $\sigma_{\eta} \geq 0$ in order to turn the non-trivial primal and dual inequalities into equalities, 
\begin{equation*}
\begin{aligned}
A_{zz}z + A_{z\eta}\eta - s_z = b_z, && A_{zz}^Ty_z + A_{\eta z}^Ty_{\eta} + \sigma_z = c_z, \\
A_{\eta z}z + \eta - s_{\eta} = b_{\eta}, && A_{z\eta}^Ty_z + y_{\eta} + \sigma_{\eta} = c_{\eta}.
\end{aligned}
\end{equation*}
However, Newton's method reduces to solving \eqref{eq:compactForm:KKTsys} with $D_1$, $D_2$, $D_3$, and $D_4$ in the coefficient matrix respectively replaced by $\Sigma_z = \diag(\sigma_z)$, $S_z = \diag(s_z)$, $\Sigma_{\eta} = \diag(\sigma_{\eta})$, and $S_{\eta} = \diag(s_{\eta})$. The computational effect on a single iteration is therefore insignificant. 

\subsection{Performance of method implementation}
We finalize this section by presenting the results of applying a MATLAB R2015a implementation of the previously described interior-point method to the cases introduced in Section~\ref{sec:CompStud}. The progress of the method is shown in Table~\ref{tab:Performance}, with the results of using the commercial software CPLEX 12 (IBM, Armonk, New York) given as a reference. 
\begin{table}[t]\centering
\caption{Iteration progress of the interior-point method (IP) described in 
Section~\ref{sec:Method:Specific} applied to the prostate and lung cases. \emph{Dim} is the size of the original system of linear equations, whereas \emph{reduced dim} and \emph{sparsity} concern the Schur complement.\label{tab:Performance}}
\begin{tabularx}{\columnwidth}{l X X X X X}
\hline\hline
\multicolumn{6}{l}{\bf Prostate case}                                 \\
\multicolumn{6}{l}{Dim 459\,638, reduced dim 3954, sparsity 11.16 \%}  \\[3pt]
Solver & Dual gap & Nb fact & Nb solv & Obj val & Residual   \\\hline
IP     & 7.04e-01 & 29      & 107     & 57.684  & 1.13e-02   \\
       & 3.45e-02 & 33      & 127     & 57.433  & 8.89e-04   \\
       & 7.47e-03 & 35      & 133     & 57.424  & 2.28e-04   \\
       & 7.76e-04 & 37      & 142     & 57.421  & 3.56e-05   \\
       & 2.81e-06 & 39      & 156     & 57.421  & 1.27e-07   \\
       & 4.72e-08 & 40      & 159     & 57.421  & 2.22e-09   \\[3pt]
CPLEX  & 4.75e-05 & 26      & ---     & 57.421  & 3.77e-15   \\[6pt] 
\multicolumn{6}{l}{\bf Lung case}                                     \\
\multicolumn{6}{l}{Dim 746\,649, reduced dim 11\,799, sparsity 11.13 \%}  \\[3pt]
Solver & Dual gap & Nb fact & Nb solv & Obj val & Residual   \\\hline
IP     & 7.10e-01 & 34      & 142     & 47.410  & 1.81e-02   \\
       & 7.36e-02 & 38      & 163     & 47.159  & 2.21e-03   \\
       & 9.50e-03 & 41      & 173     & 47.134  & 3.16e-04   \\
       & 1.36e-04 & 44      & 190     & 47.130  & 5.06e-06   \\
       & 1.94e-05 & 45      & 195     & 47.130  & 1.06e-06   \\
       & 3.29e-07 & 46      & 201     & 47.130  & 1.18e-08   \\[3pt]
CPLEX  & 1.10e-04 & 40      & ---     & 47.130  & 6.02e-06   \\
\hline\hline
\end{tabularx}
\end{table} 
The total number of factorizations (one per iteration) and solves (up to ten per iteration) is indicated in columns \emph{Nb fact} and \emph{Nb solv}. \emph{Dual gap} gives an upper bound on the gap between the current objective function value (\emph{Obj val}) and the optimum, and can be used as a stopping criterion. \emph{Residual} is the current primal and dual feasibility. 

As seen in Table~\ref{tab:Performance}, the required number of iterations was 40 and 46 to meet the accuracy criterion. On an Intel Core i7 2.80 GHz computer, computing and factorizing the Schur complement took about 4.9 and 0.46 seconds for the prostate case, resulting in a total running time of approximately 5 minutes. For the lung case, 59.5 and 9.2 seconds translate into a running time of about 60 minutes. This should be contrasted to the several hours required by CPLEX to solve each of the cases, indicating the importance of accounting for problem structure when solving \eqref{eq:convexFMO}.

\section{Discussion}
Treatment planning by conventional means is known to be a complex process involving several re-optimizations with successively fine-adjusted parameters. In this study, we have dealt with one possible cause: inconsistency between the criteria used for optimizing and evaluating treatment plans. We propose planning objectives with an explicit relationship to commonly used plan quality measures, in our case the DVH statistics, and have thereby left the usual penalty-function framework used by both the conventional and other suggested planning objectives \cite{Bokrantz2013b,Kessler2005,Romeijn2006}.

In an initial computational study involving fluence map optimization of two patient cases, we explored the potential of the proposed framework as a tool for optimizing the DVH statistics. We hypothesized that the doses-at-volume (i.e., individual points on the DVH curve) of treatment plans optimized using the proposed planning objectives would be ranked as better than the doses-at-volume of plans generated within the conventional framework. Dominance in this aspect was indeed observed in cohorts of differently balanced plans, and in addition, a larger variety of doses-at-volume was seen among plans optimized within the proposed framework. The indication is that the DVH statistics are better optimized and more efficiently balanced by the proposed planning objectives than by the conventional approach.

Despite its trial-and-error nature, it cannot be denied that the conventional planning process is able to produce treatment plans of high clinical relevance. Whether the same is true for treatment plans optimized using the proposed framework was beyond the scope of this preliminary study to explore. Examination of the clinical acceptability of treatment plans is of high importance to draw further conclusions regarding the proposed planning objectives and should be covered by future investigation, for instance by looking more closely at dose distributions and DVH curves. The analysis is even strengthened if deliverability constraints are added, which relate to the physical limitations of treatment machines. 

Clinical relevance also depends on the availability of fast optimization methods. The method described in Section~\ref{sec:Method} indicates that the proposed formulation is solvable if its structure is accounted for. The method is not as efficient as most commercial treatment planning systems, and the computational cost is expected to increase even more if deliverability constraints are added. However, given that the conventional framework requires a considerable amount of manual overhead, longer optimization running times could be acceptable if the planning process necessitates less user guidance.

\section{Conclusion}
We have formulated planning objectives for treatment plan multicriteria optimization with an explicit relationship to DVH statistics. This is in contrast to the conventionally clinically used planning objectives by which the violation of DVH statistics thresholds is minimized, offering limited control of individual points on the DVH curve (doses-at-volume). The merit of the two planning approaches as tools for DVH statistics optimization was investigated by exploring sets of differently balanced treatment plans generated using each approach. Dominance was observed, in the sense of better doses-at-volume, among the sets of plans optimized within the proposed framework. In addition, a larger variety of doses-at-volume was seen in these sets of plans, indicating that the DVH statistics are better optimized and more efficiently balanced using the proposed planning objectives. 

Treatment planning with the proposed planning objectives amounts to solving optimization problems whose dimensions are of the same order as the generally large number of voxels. Availability of a numerical method that can handle these large problems is of utmost importance for the validity of the proposed framework. We have demonstrated how the problem structure can be used to adapt a standard optimization method, and thereby obtain a reduction in computational cost by several orders of magnitude. 

\bibliography{convfmo_abbr}
\bibliographystyle{siam} 
\end{document}